\documentclass[12pt]{article}
\usepackage{amsmath,amsfonts,amssymb}
\usepackage{hyperref}
\usepackage{graphicx}
\usepackage[dvips]{color}
\usepackage{epsfig}
\usepackage{psfrag}
\makeatletter\@addtoreset{equation}{section}\makeatother

\def\bR {\mathbb{R}}

\newcommand{\beq}{\begin{equation}}
\newcommand{\eeq}{\end{equation}}
\newcommand{\bal}{\begin{equation}\begin{aligned}}
\newcommand{\eal}{\end{aligned}\end{equation}}
\newcommand{\bea}{\begin{eqnarray}}
\newcommand{\eea}{\end{eqnarray}}

\newcommand{\vev}[1]{{\left< {#1} \right>}}
\newcommand{\bra}[1]{{\left< {#1} \right|}}
\newcommand{\ket}[1]{{\left| {#1} \right>}}

\newcommand{\eqn}[1]{(\ref{#1})}

\newcommand{\address}[1]{\vbox{\center\em#1}}
\renewcommand{\title}[1]{\vbox{\center\LARGE{#1}}\vspace{5mm}}

\newcommand{\cF}{{\mathcal F}}

\newcommand{\cN}{{\mathcal N}}

\newcommand{\cI}{{\mathcal I}}

\begin{document}
\bibliographystyle{utphys}

\begin{titlepage}
\begin{center}
\vspace{5mm}
\hfill {\tt HU-EP-10/70}\\
\hfill {\tt Imperial-TP-2010-ND-01}\\
\hfill {\tt NSF-KITP-10-155}\\
\vspace{14mm}

\title{(de)Tails of Toda CFT}
\vspace{15mm}

\renewcommand{\thefootnote}{$\alph{footnote}$}

Nadav Drukker\footnote{\href{mailto:n.drukker@imperial.ac.uk}
{\tt n.drukker@imperial.ac.uk}}
and
Filippo Passerini\footnote{\href{mailto:filippo@physik.hu-berlin.de}
{\tt filippo@physik.hu-berlin.de}}
\vskip 5mm
\address{
${}^{a}$The Blackett Laboratory, Imperial College London,\\
Prince Consort Road, London SW7 2AZ, U.K.\\
}
\address{
${}^{b}$Institut f\"ur Physik, Humboldt-Universit\"at zu Berlin,\\
Newtonstra\ss e 15, D-12489 Berlin, Germany
}

\renewcommand{\thefootnote}{\arabic{footnote}}
\setcounter{footnote}{0}

\end{center}

\vspace{8mm}
\abstract{
\normalsize{
\noindent
The relation between the partition function of $\cN=2$ gauge theories in 4d and 
conformal Toda field theory in 2d is explained for the case where the 4d theory 
is a linear quiver with ``quiver tails''. That is when the 4d theory has gauge groups 
of different rank. We propose an identification of a subset of the states of Toda CFT 
which represent the Coulomb-branch parameters of the different rank gauge multiplets 
and study their three-point functions and descendants.
}}
\vfill

\end{titlepage}

\section{Introduction}
\label{sec:intro}

The Alday-Gaiotto-Tachikawa (AGT) relation \cite{AGT} expresses the partition function of certain 
4d gauge theories with $\cN=2$ supersymmetry on $S^4$ in terms of correlation functions 
in 2d Liouville theory. More precisely, such 4d theories are conjectured to have S-duality 
symmetries, and the identification is between the partition function calculated in one 
S-duality frame and the Liouville correlation function calculated via the conformal bootstrap 
with a particular pants decomposition of the Riemann surface.%
\footnote{Some refer to the AGT relation as the identification between instanton 
partition functions and Liouville conformal blocks, which is also true.}

The gauge theory partition function on $S^4$ is given by an integral over Coulomb branch 
parameters for each of the gauge groups with certain classical, 1-loop, and instanton contributions 
\cite{Pestun:2007rz}. The identification is between:
\begin{table}[h]
\begin{center}
\begin{tabular}{|p{2.2in}||p{3in}|}
\hline
Gauge theory& Liouiville\\
\hline 
& \\
[-1.2em]
\hline
Coulomb branch parameters&Primary states in intermediate channels\\
\hline
1-loop corrections&Three point functions\\
\hline
Instanton contributions&Contributions of descendants, captured by the conformal blocks.\\
\hline
\end{tabular}
\end{center}
\end{table}

The gauge theories related to Liouville theory in this way have gauge group $SU(2)^k$ for some $k$. 
The AGT relation was generalized by Wyllard \cite{Wyllard:2009hg} to the case of other gauge 
groups, namely $SU(N)^k$. The identification goes as before, with Liouville theory replaced by 
$A_{N-1}$ Toda CFT \cite{Zamolodchikov:1985wn, Fateev:1985mm}.

We would like to address a hybrid case, when the gauge group is a product $\prod_iSU(N_i)$ with 
different values of $N_i$. We will consider only the case when the theory has a Lagrangean in 
at least one duality frame, so we exclude the theories with $T_N$ factors \cite{Gaiotto:2009we}, 
and when the 
theories are conformal (apart for explicit mass terms). The classification of these 
theories was done by by Gaiotto and Witten \cite{Gaiotto:2009we, Gaiotto:2008ak} 
and involves a linear quiver of gauge groups 
$SU(N)$ with extra quiver tails of lower rank gauge groups, as we review below.

We would like to reproduce the partition function of such a theory from a correlation function 
of $A_{N-1}$ Toda CFT. Since the AGT relation identifies the coulomb branch parameters 
with primary states in intermediate channels along the surface, for groups of rank smaller 
than $N$ in the tail, the space of states in the corresponding intermediate channel should be 
smaller than the full $N-1$ dimensional space of Toda primaries. The purpose of this 
manuscript is to propose what this reduced space of states is, so that we can generalize 
the AGT correspondence to this case too.

In the next section we review the classification of quiver tails and its relation to semi-degenerate 
representations of Toda CFT \cite{KMST}. A careful analysis of the four-dimensional theory leads to a 
guess on the allowed intermediate states in the relevant channels, which is backed by an analysis 
of the Seiberg-Witten curve. In Section~\ref{sec:2x3} 
we study the simplest theory of this type, with gauge group $SU(2)\times SU(3)$. In that 
case one can write down a Ward identity for the 3-point function \cite{KMS} whose solution 
is exactly the space of states we proposed. The three point function and its relation to the 
1-loop determinant is discussed in much more generality in the following sections. 
So here we concentrate on the contribution of level 1 descendants and show that it 
agrees with the instanton 
correction to the gauge theory expression. We note, in particular that the sum over 
W-algebra descendants of the restricted state representing $SU(2)$ reproduces the 
same instanton contribution as do the Virasoro descendants in Liouville CFT. In this 
case we also write down the precise Seiberg-Witten curve and match the mass parameters 
and intermediate states to a semiclassical limit of the Toda states.

In Section~\ref{sec:3pt} we evaluate the relevant 3-point functions. 
We start with the expression for the 3-point function of a ``simple puncture'' state 
and two generic (``full puncture'') states \cite{Fateev:2005gs, Fateev:2007ab} and consider the limit when these two 
states become semi-degenerates. We show that in that limit the 3-point function will 
generically vanish, except for the special states that we guessed in Section~\ref{sec:tails}, 
where we find a pole of the expected order.

In Section~\ref{sec:together} we combine the expressions we found for the 3-point functions 
for the entire tail and show how they reproduce the correct 1-loop part in the gauge theory 
calculation.

We discuss some further issues in Section~\ref{sec:discuss}.

\section{Quiver tails and semi-degenerate representations}
\label{sec:tails}

\subsection{Classification of quiver tails}
\label{sec:classify-tails}

Let us recall here the classification of conformal $SU(N)$ linear quiver tails 
\cite{Gaiotto:2009we, Gaiotto:2008ak}.

An $\cN=2$ supersymmetric $SU(N)$ gauge theory is conformal when the total 
number of matter fields in the fundamental and anti-fundamental representations 
is $N_F=2N$. Any subgroup 
of the flavor symmetry may be gauged, but then we should ask whether 
the resulting gauge theory is still conformal. The simplest theories constructed in this 
way have some number $k$ of $SU(N)$ gauge multiplets arranged along a line 
with nearest neighbors having bi-fundamental matter charged under the two groups. 
At the end of this linear quiver we can add bi-fundamental matter between the first 
and last group, getting a quiver ring. It is also possible to add $N$ fundamental fields to the first and 
$N$ anti-fundamental to the last, giving a linear quiver.

In the latter configuration, at each end of the linear quiver there is an $SU(N)$ flavor symmetry, and we 
can try to gauge a subgroup of it. For example, if $N$ is even we can gauge an $SU(N/2)$ 
subgroup. This gauge theory will be conformal, since there are $N$ fields in the fundamental of 
$SU(N/2)$. Such constructions can be done at either end of the quiver and involve 
a finite series of groups of decreasing rank. The most general linear quiver tail 
\cite{Gaiotto:2009we, Gaiotto:2008ak} has gauge group
\beq
SU(N_1)\times SU(N_2)\times\cdots\times SU(N_{k-1})\times SU(N_k)\,,
\qquad N_k=N\,.
\eeq
The requirement of conformality is that the sequence of integers $N_1,\cdots, N_k$ is 
convex, {\em i.e.}
\beq
N_1\geq N_2-N_1\geq N_3-N_2\geq\cdots\geq N_k-N_{k-1}\,.
\eeq
The $SU(N_i)$ group couples to $N_{i-1}+N_{i+1}$ fields%
\footnote{Here $N_0=0$.} 
from the two bi-fundamentals so that by adding $2N_i-(N_{i-1}+N_{i+1})\geq0$ extra fundamental 
fields, we get a conformal theory.

\begin{figure}[!ht]
\begin{center}
\vskip4mm
\epsfig{file= 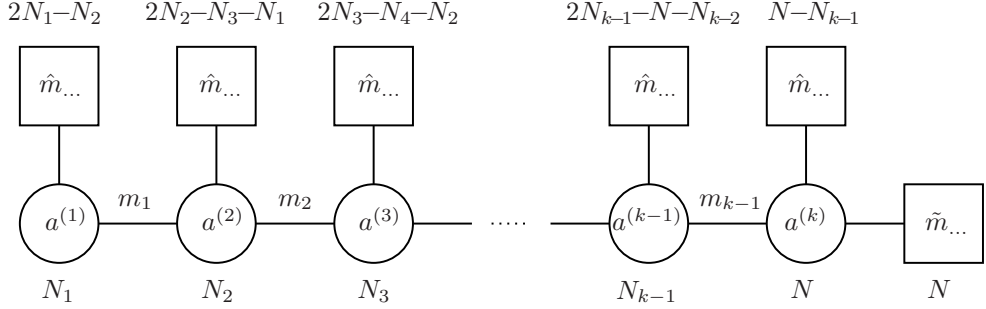,width=130mm
\psfrag{N1}{\footnotesize $N_1$}
\psfrag{N2}{\footnotesize $N_2$}
\psfrag{N3}{\footnotesize $N_3$}
\psfrag{Nk-1}{\footnotesize $N_{k-1}$}
\psfrag{N}{\footnotesize $N$}
\psfrag{a1}{\footnotesize $a^{(1)}$}
\psfrag{a2}{\footnotesize $a^{(2)}$}
\psfrag{a3}{\footnotesize $a^{(3)}$}
\psfrag{ak-1}{\footnotesize $a^{(k-1)}$}
\psfrag{ak}{\footnotesize $a^{(k)}$}
\psfrag{ak+1}{\footnotesize $\ \ \tilde m_{\ldots}$}
\psfrag{a0}{\footnotesize $\hat m_{\dots}$}
\psfrag{m2}{\footnotesize $m_{1}$}
\psfrag{m3}{\footnotesize $m_{2}$}
\psfrag{mk}{\footnotesize $\!\!m_{k-1}$}
\psfrag{2N1-N2}{\footnotesize $2N_1\!\!-\!\!N_2$}
\psfrag{2N2-N3-N1}{\footnotesize $2N_2\!\!-\!\!N_3\!\!-\!\!N_1$}
\psfrag{2N3-N4-N2}{\footnotesize $2N_3\!\!-\!\!N_4\!\!-\!\!N_2$}
\psfrag{2Nk-1-N-Nk-2}{\footnotesize $2N_{k\!-\!1}\!\!-\!\!N\!\!-\!\!N_{k\!-\!2}$}
\psfrag{N-Nk-1}{\footnotesize $N\!\!-\!\!N_{k\!-\!1}$}
} 
\parbox{13cm}{
\caption{
A general quiver tail. We explain our notations for the fundamental matter $\hat m_{\cdots}$ in the text.
\label{fig:tail}}}
\end{center}
\end{figure}

The most general linear quiver tail is illustrated in Figure~\ref{fig:tail}. The circles are gauge 
multiplets, the boxes are fundamental fields and the horizontal lines are bi-fundamentals. 
On the right this quiver is terminated on $N$ anti-fundamentals, but it can also be continued 
with more $SU(N)$ gauge fields and another quiver tail. The flavor symmetry is 
given by the product of $U(2N_i-N_{i-1}+N_{i+1}))$, an extra $U(N)$ from the fundamental 
field on the right and $k-1$ factors of $U(1)$. We will also add masses to 
the fields, where each of the fundamental fields can have a mass and likewise the 
bi-fundamentals. There are in total $\sum_i(2N_i-N_{i-1}+N_{i+1}))=N_1$ mass 
parameters for the fundamental fields and $k-1$ mass parameters for the 
bi-fundamentals. In our conventions all masses as well as Coulomb branch 
parameters are purely imaginary.

We can encode the structure of the quiver tail in a Young-diagram with $N$ boxes by 
taking the $r^\text{th}$ row to be of length $N_l-N_{l-1}$. It will be useful also to 
look at the {\em columns} in the diagram and we label the Young-diagram by the sequence of 
heights $[n_1,n_2,\cdots,n_{N_1}]$. 

The longest quiver tail has $N_1=2$ and $N_l-N_{l-1}=1$ for all $l>1$ and has 
$N-1$ rows. It has two columns with $[N-1,1]$ and is also referred to as a simple 
puncture. The shortest quiver has $k=1$, so it's a single row, 
or $[1,1,\cdots,1]$, also known as a full puncture.

\begin{figure}[!ht]
\begin{center}
\epsfig{file= 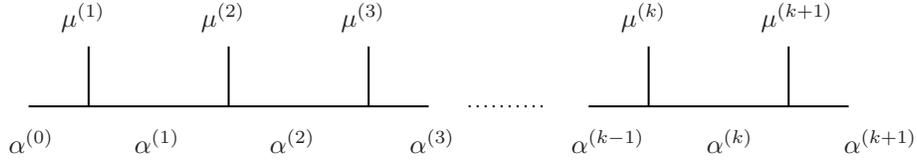,width=120mm
\psfrag{a0}{\footnotesize $\alpha^{(0)}$}
\psfrag{a1}{\footnotesize $\alpha^{(1)}$}
\psfrag{a2}{\footnotesize $\alpha^{(2)}$}
\psfrag{a3}{\footnotesize $\alpha^{(3)}$}
\psfrag{ak-1}{\footnotesize $\alpha^{(k-1)}$}
\psfrag{ak}{\footnotesize $\alpha^{(k)}$}
\psfrag{ak+1}{\footnotesize $\alpha^{(k+1)}$}
\psfrag{m1}{\footnotesize $\mu^{(1)}$}
\psfrag{m2}{\footnotesize $\mu^{(2)}$}
\psfrag{m3}{\footnotesize $\mu^{(3)}$}
\psfrag{mk}{\footnotesize $\mu^{(k)}$}
\psfrag{mk+1}{\footnotesize $\mu^{(k+1)}$}
} 
\parbox{13cm}{
\caption{
The corresponding trivalent graph.
\label{fig:graph}}}
\end{center}
\end{figure}

\subsection{Semi-degenerate representations of Toda CFT}
\label{sec:semi-deg}

The classification of quiver tails in terms of Young-diagrams matches the classification of 
physical semi-degenerate representations of conformal Toda theory \cite{KMST}. 
Semi-degenerate fields have descendants which are null vectors, the restriction to 
{\em physical} semi-degenerates is that all the null states are at level 1 (or are 
descendants of other null states). If the state satisfies $N-1$ independent null 
conditions the state is completely degenerate, and if they are all at level 1, it's the 
identity state.

A generic primary state of $A_{N-1}$ Toda CFT is given by $N-1$ continuous 
parameters (modulo $S_N$ Weyl reflections). It can be easily written in the 
orthonormal basis in terms of an $N$-vector whose components sum to zero. 
Due to the constraints satisfied by the semi-degenerate states, they will have 
fewer continuous parameters.

To be specific, the vertex operator for a primary state of Toda CFT is given (ignoring 
normal ordering issues) by
\beq
e^{\vev{\alpha,\phi}}\,,
\eeq
where $\phi$ is the Toda field defined on the root space of the 
$A_{N-1}$ algebra and the vector $\alpha$ has the form
\beq
\alpha=\vec Q+\gamma\,,
\eeq
where $\vec Q$ is the background charge%
\footnote{While many other quantities are vectors, we write the arrow only over 
$Q$ to distinguish the vector $Q\rho$ from the constant $Q$.}
\beq
\vec Q=Q\rho\,,\qquad
Q=b+\frac{1}{b}\,,
\eeq
and $\rho$ is the Weyl vector of $A_{N-1}$, that is half the sum of all the positive roots. 
$b$ is the coupling constant of the action of Toda. 
The Virasoro central charge is
\beq
c=N-1+Q^2N(N^2-1)\,.
\eeq

For a generic state of Toda, $\gamma$ is purely imaginary and breaks the 
entire $S_N$ Weyl group. For the physical (level 1) degenerate fields, the weight 
$\alpha$ is invariant under a subgroup of the Weyl group.%
\footnote{For more details on semi-degenerate states see \cite{KMST,DGG}.}
Since $\vec Q$ is real and 
breaks the Weyl group, $\gamma$ is not purely imaginary anymore.
To see the relation to Young-diagrams, for a state where $\alpha$ breaks $S_N$ to 
$S_{n_1}\times S_{n_2}\times\cdots\times S_{n_{N_1}}$ we associate the Young-diagram 
$[n_1,n_2,\cdots, n_{N_1}]$.%
\footnote{$S_1$ is of course trivial. We included them to get the sum $\sum n_i=N$.}
Such a state will depend on $N_1-1$ continuous imaginary parameters, one for 
each column in the diagram, with the weighted sum vanishing. We write explicit expressions 
for such states below in \eqn{alpha0}.

\subsection{Subspaces of primaries of Toda CFT}
\label{sec:subspaces}

\begin{figure}[!t]
\begin{center}
\vskip4mm
\epsfig{file= 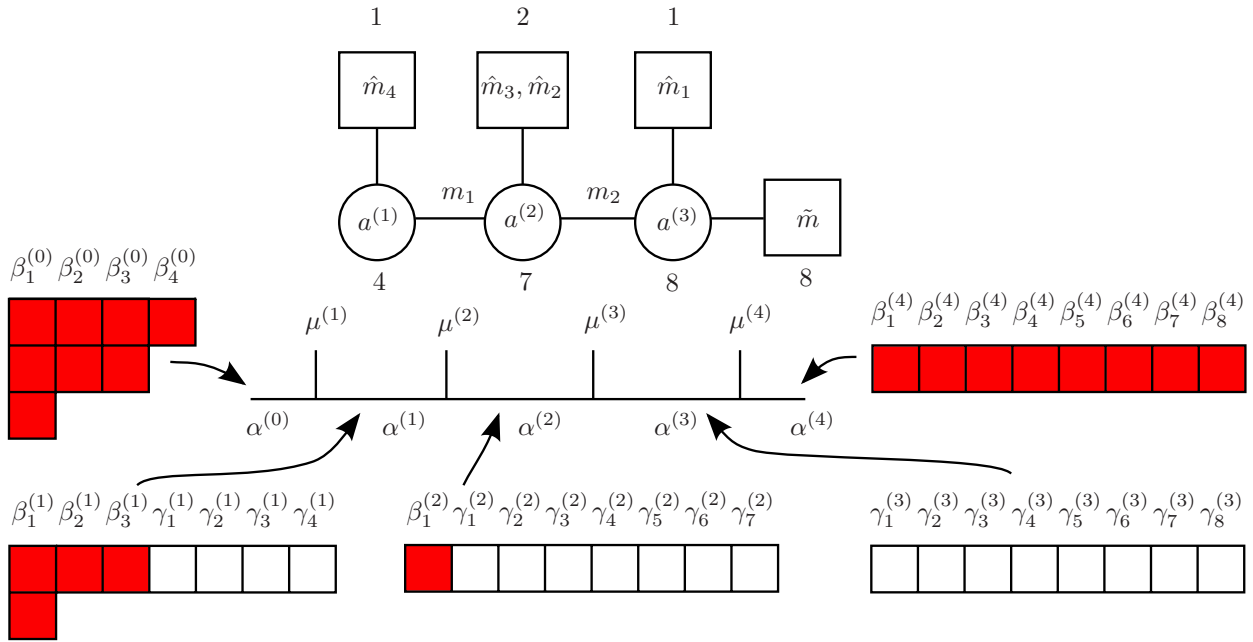,width=165mm
\psfrag{1}{\footnotesize$1$}
\psfrag{2}{\footnotesize$2$}
\psfrag{4}{\footnotesize$4$}
\psfrag{7}{\footnotesize$7$}
\psfrag{8}{\footnotesize$8$}
\psfrag{mh1}{\footnotesize $\hat m_1$}
\psfrag{mh2}{\footnotesize $\hat m_3,\hat m_2$}
\psfrag{mh4}{\footnotesize $\hat m_4$}
\psfrag{a1}{\footnotesize $a^{(1)}$}
\psfrag{a2}{\footnotesize $a^{(2)}$}
\psfrag{a3}{\footnotesize $a^{(3)}$}
\psfrag{a4}{\footnotesize $\ \tilde m$}
\psfrag{m1}{\footnotesize $m_1$}
\psfrag{m2}{\footnotesize $m_2$}
\psfrag{mu1}{\footnotesize $\mu^{(1)}$}
\psfrag{mu2}{\footnotesize $\mu^{(2)}$}
\psfrag{mu3}{\footnotesize $\mu^{(3)}$}
\psfrag{mu4}{\footnotesize $\mu^{(4)}$}
\psfrag{al0}{\footnotesize $\alpha^{(0)}$}
\psfrag{al1}{\footnotesize $\alpha^{(1)}$}
\psfrag{al2}{\footnotesize $\alpha^{(2)}$}
\psfrag{al3}{\footnotesize $\alpha^{(3)}$}
\psfrag{al4}{\footnotesize $\alpha^{(4)}$}
\psfrag{b01}{\footnotesize $\beta^{(0)}_1$}
\psfrag{b02}{\footnotesize $\beta^{(0)}_2$}
\psfrag{b03}{\footnotesize $\beta^{(0)}_3$}
\psfrag{b04}{\footnotesize $\beta^{(0)}_4$}
\psfrag{b11}{\footnotesize $\beta^{(1)}_1$}
\psfrag{b12}{\footnotesize $\beta^{(1)}_2$}
\psfrag{b13}{\footnotesize $\beta^{(1)}_3$}
\psfrag{g11}{\footnotesize $\gamma^{(1)}_1$}
\psfrag{g12}{\footnotesize $\gamma^{(1)}_2$}
\psfrag{g13}{\footnotesize $\gamma^{(1)}_3$}
\psfrag{g14}{\footnotesize $\gamma^{(1)}_4$}
\psfrag{b21}{\footnotesize $\beta^{(2)}_1$}
\psfrag{g21}{\footnotesize $\gamma^{(2)}_1$}
\psfrag{g22}{\footnotesize $\gamma^{(2)}_2$}
\psfrag{g23}{\footnotesize $\gamma^{(2)}_3$}
\psfrag{g24}{\footnotesize $\gamma^{(2)}_4$}
\psfrag{g25}{\footnotesize $\gamma^{(2)}_5$}
\psfrag{g26}{\footnotesize $\gamma^{(2)}_6$}
\psfrag{g27}{\footnotesize $\gamma^{(2)}_7$}
\psfrag{g31}{\footnotesize $\gamma^{(3)}_1$}
\psfrag{g32}{\footnotesize $\gamma^{(3)}_2$}
\psfrag{g33}{\footnotesize $\gamma^{(3)}_3$}
\psfrag{g34}{\footnotesize $\gamma^{(3)}_4$}
\psfrag{g35}{\footnotesize $\gamma^{(3)}_5$}
\psfrag{g36}{\footnotesize $\gamma^{(3)}_6$}
\psfrag{g37}{\footnotesize $\gamma^{(3)}_7$}
\psfrag{g38}{\footnotesize $\gamma^{(3)}_8$}
\psfrag{g41}{\footnotesize $\beta^{(4)}_1$}
\psfrag{g42}{\footnotesize $\beta^{(4)}_2$}
\psfrag{g43}{\footnotesize $\beta^{(4)}_3$}
\psfrag{g44}{\footnotesize $\beta^{(4)}_4$}
\psfrag{g45}{\footnotesize $\beta^{(4)}_5$}
\psfrag{g46}{\footnotesize $\beta^{(4)}_6$}
\psfrag{g47}{\footnotesize $\beta^{(4)}_7$}
\psfrag{g48}{\footnotesize $\beta^{(4)}_8$}
} 
\parbox{15cm}{
\caption{An example of the AGTails correspondence. The gauge group is 
$SU(4)\times SU(7)\times SU(8)$ and the parametrization of the 
Toda states follows \eqn{alpha0} and \eqn{alphai}. 
Successive Young-diagrams are given by merging the 
first two rows such that the white boxes are those which were on the first line of the 
previous diagram. The parameters $\beta^{(l)}_i$, related to the 
red boxes are fixed by the mass parameters in the 
gauge theory. The $\gamma^{(l)}_i$ parameters at the white boxes 
are related to the Coulomb branch parameters in the gauge theory. 
\label{fig:example}}}
\end{center}
\end{figure}

According to the AGT correspondence we should associate to each gauge theory 
a Riemann surface with punctures, decorated by mass parameters. In the case of linear 
quivers, the surface is a punctured sphere. The number of punctures is the number of 
gauge group components plus three.%
\footnote{A simple way to do the counting it to recall that the complex moduli 
parameters of the surface are mapped to the gauge couplings.}

Since the classification of quiver tails matches the physical semi-degenerate fields of 
Toda, it is natural to expect \cite{KMST} that for each of the tails there will be one puncture with 
the semi-degenerate state. Indeed the analysis of the Seiberg-Witten curve for 
theories with a quiver tail gives a special singular point, and many other regular singularities 
(``simple punctures''). In the case of the theory in Figure~\ref{fig:tail} there 
are $k+1$ simple punctures, one full puncture, corresponding to the right-most matter 
in the quiver and a special puncture where we will insert the semi-degenerate state.

We expect to be able to reproduce the partition function of this gauge theory by studying 
Toda CFT on this Riemann surface. We consider the trivalent graph in 
Figure~\ref{fig:graph} representing the successive fusion of the semi-degenerate 
state $\alpha^{(0)}$ with simple-puncture states $\mu^{(l)}$. The intermediate states 
$\alpha^{(l)}$ with $1\leq l\leq k$ have to be integrated over. They should be related 
to the Coulomb branch parameters $a^{(l)}$, which take values in $i\bR^{N_l-1}/S_{N_l}$, 
and therefore the states $\alpha^{(l)}$ should be restricted to such a subspace of 
primaries of Toda.

We can now state in a precise way the question we would like to answer:
\begin{quote}
{\em
\begin{enumerate}
\item
How are the mass parameters $m_l$ and $\hat m_l$ in the gauge theory 
encoded in the states $\mu^{(l)}$, $\alpha^{(0)}$ and $\alpha^{(k+1)}$.
\item
What are the allowed states $\alpha^{(l)}$ in the intermediate channels and how 
are they related to the Coulomb branch parameters $a^{(l)}$.
\end{enumerate}
}
\end{quote}

\subsubsection*{\rm Our proposal is the following:}
\begin{itemize}
\item
The mass parameters determine $\alpha^{(0)}$, $\mu^{(l)}$ and $\alpha^{(k+1)}$ according 
to \eqn{beta} and \eqn{initial}.
\item
The intermediate states $\alpha^{(l)}$ are semi-degenerate states. Their Young-diagrams is 
gotten from that of $\alpha^{(l-1)}$ by combining the first two rows into a single row.
\item
The Coulomb branch parameters $a^{(l)}$ are encoded in a subset of the directions of the vector 
$\alpha^{(l)}$ which are not degenerate. The other directions are related to the 
mass parameters to the right of that state, see \eqn{beta} and \eqn{initial}.
\end{itemize}

To start we should specify how we encode the mass parameters of the fundamental 
fields in $\hat m$. There are in total $N_1$ fundamental fields for the different groups 
(excluding the $N$ fundamentals on the right), one for each column in the Young-diagram. 
We label $\hat m_1$ the mass of the right most fundamental field in the quiver and 
$\hat m_{N_1}$ the left most.

The state $\alpha^{(0)}$ has $N_1-1$ continuous parameters and a symmetry 
$S_{n_1}\times\cdots\times S_{N_{N_1}}$. We may 
parametrize $\alpha^{(0)}$ in the orthonormal basis as
\bal
\label{alpha0}
\alpha^{(0)}=\vec Q+\big(
&
\beta^{(0)}_{N_1}+\delta_{n_{N_1},1},\cdots,\beta^{(0)}_{N_1}+\delta_{n_{N_1},n_{N_1}},
\quad \cdots\quad ,
\beta^{(0)}_1+\delta_{n_1,1},\cdots,\beta^{(0)}_1+\delta_{n_1,n_1}),
\eal
where
\beq
\sum_{j=1}^{N_1}n_j\beta^{(0)}_j=0\,.
\eeq
The parameters $\delta_{n,j}=(2j-n-1)Q/2$ are necessary to make 
the state $\alpha^{(0)}$ invariant under the desired subgroup of $S_N$.%
\footnote{In \cite{DGG} these terms were grouped together into $Q\rho_\cI$.}

The Young-diagram has $2N_1-N_2$ columns of unit height (for which $n_i=1$ and 
$\delta_{1,1}=0$). We write the $\beta^{(0)}_i$ parameters associated to them 
at the left of the vector, with the parameters for the other columns of the 
Young-diagram further on the right.

Our claim is that the state $\alpha^{(l)}$ will have a Young-diagram with rows of length 
$N_{l+1}, N_{l+2}-N_{l+1},\cdots,N-N_{k-1}$. There are $2N_{l+1}-N_{l+2}$ columns 
of unit height and the others are of height $n_j-l$, where $n_j$ are the heights of the 
columns in the original Young-diagram (including only the columns for which $n_j-l>1$).

We parameterize the intermediate states of this type in the orthonormal basis as
\bal
\label{alphai}
\alpha^{(l)}=\vec Q+\big(
& \bar\beta^{(l)}+\gamma^{(l)}_1,\cdots, \bar\beta^{(l)}+\gamma^{(l)}_{N_l},
\\&
\beta^{(l)}_{N_{l+1}-N_{l}}+\delta_{n_{N_{l+1}-N_{l}}-l,1},\cdots,
\beta^{(l)}_{N_{l+1}-N_{l}}+\delta_{n_{N_{l+1}-N_{l}}-l,n_{N_{l+1}-N_{l}}-l},
\\&\cdots,
\\&\beta^{(l)}_1+\delta_{n_1-l,1},\cdots,\beta^{(l)}_1+\delta_{n_1-l,n_1-l}),
\eal
where
\beq
N_l\bar\beta^{(l)}=-\sum_{i=1}^{N_{l+1}-N_l}(n_i-l)\beta^{(l)}_i\,,
\qquad\text{and}\qquad
\sum_{i=1}^{N_l}\gamma^{(l)}_i=0\,.
\eeq
The simple puncture states $\mu^{(l)}$ are also semi-degenerate with 
Young-diagrams with columns $[N-1,1]$ and one continuous parameter $\mu_l$. 
We take them to be proportional to the first fundamental weight 
$\mu^{(l)}=\big({\textstyle\frac{Q}{2}}-\mu_l\big)N\omega_1$.

The identification we propose is that $\gamma^{(l)}$ are the $N_l-1$ 
Coulomb branch parameters $a^{(l)}_i$ of $SU(N_l)$. The parameters 
$\beta^{(l)}_i$ as well as the parameters of the simple punctures are given by the 
mass parameters as the solution to the following set of linear equations
\bal
\label{beta}
\mu_l-\beta^{(l-1)}_i+\beta^{(l)}_i&=0\,,
\qquad
&l&=1,\cdots,n_i\,,
\\
\mu_l-\bar\beta^{(l-1)}+\bar\beta^{(l)}&=m_{l-1}\,,
\qquad
&l&=2,\cdots,n_1\,.
\eal
with the initial values%
\footnote{$\beta^{(n_i)}_i$ with these indices do not appear in the 
parametrization of the states above, but are useful for writing the equations.}
\beq
\label{initial}
\beta^{(n_i)}_i=\bar\beta^{(n_i)}+\hat m_i\,,
\qquad
\bar\beta^{(n_1)}=0\,.
\eeq
These equations can be solved recursively from large to small $l$.

The examples we concentrate on in the following sections are of Toda theory with 
$N+1$ simple punctures and one full puncture, so that $\alpha^{(0)}$ also has a 
Young-diagram with columns $[N-1,1]$. Then the solution to the conditions \eqn{beta} is
\beq
\beta^{(l)}_1=\frac{l+1}{N}\left(\hat m_1+\sum_{i=l}^{N-2} m_i\right),
\qquad
N\mu_l=l\,m_{l-1}-\hat m_1-\sum_{i=l}^{N-2}m_i\,,
\qquad
m_0=-\hat m_2\,.
\eeq

In the rest of this manuscript we test this identification by studying the fusion rules 
of two semi-degenerate states and a simple puncture. We also compare the product of three 
point functions appearing in the tail with the 1-loop partition function arising from integrating 
out the fundamental and bi-fundamental fields in the gauge theory.

\subsection{The Seiberg-Witten curve}
\label{sec:SW}

As a first test, though, we would like to check the consistency of the construction under 
decoupling of one of the gauge groups. We consider sending to zero the coupling of the first 
gauge multiplet, that of $SU(N_1)$. In terms of the Riemann surface this is a degeneration 
limit where the neck connecting the first two punctures (with insertions $\alpha^{(0)}$ and 
$\mu^{(1)}$) is stretched to infinite length. In the limit this neck will look like a single puncture 
on the sphere with the state $\alpha^{(1)}$ and the same $k+1$ other operators as before.

On the gauge theory side we have the gauge group $SU(N_2)\times\cdots\times SU(N_k)$. 
The Coulomb branch parameters of $SU(N_1)$ and $m_1$ are now frozen as mass 
parameters for $SU(N_2)$. Looking at how we defined $\alpha^{(0)}$ and $\mu^{(1)}$ 
before exactly matches the definition of $\alpha^{(1)}$ and $\mu^{(2)}$ in terms of the 
new masses. The identification of all the other parameters $\alpha^{(l)}$ and 
$\mu^{(i+1)}$ for $i>1$ is not affected by the decoupling of the first $SU(N_1)$, 
since their definitions depended only on parameters that appear in the quiver to the 
right of $SU(N_2)$.

This can be done in more detail in the semiclassical approximation (ignoring $b$ and $b^{-1}$) 
by examining the Seiberg-Witten curve for these theories, which was written down in 
\cite{Witten:1997sc,KMST}
\begin{align}
0=&\,
\prod_{i=1}^{N}\Big(v-\tilde M_i\Big)z^{k+1}
+c_{k}\Big(v^N-M_k v^{N-1}-u^{(k)}_{2}v^{N-2}-\cdots-u^{(k)}_{N-1}v-u^{(k)}_{N}\Big)z^k +\cdots 
\nonumber\\
&\,
+c_j\Bigg(\prod_{i=1}^{N_{j+1}-N_{j}}(v-\hat{M}_i)^{n_i-j}\Bigg) \Big(v^{N_j}-M_jv^{N_j-1}
-\cdots-u^{(j)}_{N_j-1}v-u^{(j)}_{N_j}\Big)z^j
+\cdots 
\nonumber\\
&\,+c_0\prod_{i=1}^{N_1}(v-\hat{M}_i)^{n_i}
\label{witt}
\end{align}
where $c_j$ and $u^{(j)}_{s}$ with $s=2,\dots,N_j$ are the gauge coupling and a set of 
Coulomb branch parameters for the $j$-th gauge group. 
$\tilde M_i$ are the masses of the $N$ fundamental hypermultiplets coupling to the 
rightmost gauge group. $\hat{M}_i$ and $M_i$ are related to the masses of the fundamental 
hypermultiplets of the  tail and the bifundamental matter.%
\footnote{The physical masses of BPS states can be read from the poles in the 
Seiberg-Witten differential. The simplest (and most commonly cited) choice 
for the differential, which is $\lambda=v\,dz/z$, gives the correct physical masses only after 
a complicated linear transformation on the parameters $\hat M_i$ and $M_i$. This is 
explained for the case of $SU(2)$ with $N_F=4$ in \cite{Argyres:1995wt} (see also 
the discussion in Appendix~C of \cite{Hollands:2010xa}). We perform this transformation 
in detail for the case of $SU(2)\times SU(3)$ in Section~\ref{sec:SW2x3}.\label{massnote}}

This curve was rewritten by Gaiotto in a different form \cite{Gaiotto:2009we}. 
First we collect all the terms with definite power of $v$
\beq
\Delta (z)v^N + M(z)v^{N-1}+R_{N-2}(z)v^{N-2} +\cdots+R_{N-1}(z)=0
\eeq
and then shift $v$ as $v\rightarrow v-\frac{M(z)}{N\Delta(z)}$. Considering the change of 
coordinate $v=x z$ the curve can be written as 
\beq
\label{gaio}
x^N +\Phi^{(N-1)}(z)x^{N-2}+\cdots+\Phi^{(N)}(z)=0\,.
\eeq
In this way the Seiberg-Witten curve is written as an $N^\text{th}$ cover of a 
Reimann surface which in this case is the $k+3$ punctured sphere. 

The functions $\Phi^{(I)}(z)$ have poles of order $I$ at the location of each of the 
punctures $z=z_i$ \cite{Gaiotto:2009we}
\beq
\Phi^{(I)}(z)=\frac{\phi^{(I)}_{i,0}}{(z-z_i)^I}+\frac{\phi^{(I)}_{i,-1}}{(z-z_i)^{I-1}}
+\cdots
\eeq
and it was shown in \cite{AGT,KMST} that in the semiclassical approximation 
the coefficient of the leading poles are the charges under the W-symmetry of the 
Toda state inserted at the puncture
\beq
\label{agtw}
\bar{\phi}^{(I)}_{i,0}\sim w^{I}_i\,.
\eeq
It is straight-forward to read the parameters of the Toda states from 
the Seiberg-Witten curve, which is written explicitly in \eqn{witt}. One does has to 
take into account the shift in $v$ needed to write \eqn{gaio}, which act by linear 
transformations on the mass parameters. Furthermore, as mentioned in 
footnote~\ref{massnote}, these parameters are not exactly the physical masses in 
the theory. In Section~\ref{sec:SW2x3} we do it in detail for the $SU(2)\times SU(3)$ 
theory. Here we study the allowed intermediate states in the general curve, but due to 
this difficulty, we discuss only the type of states, not the exact values of the 
mass parameters.

Looking at \eqn{witt} and expanding around $z\sim0$ we can identify in the semiclassical 
approximation \cite{KMST}
\beq
\alpha^{(0)}\sim \Big(
\underbrace{\hat{M}_{N_1}^{(0)},\cdots, \hat{M}_{N_1}^{(0)}}_{n_{N_1}},
\cdots, 
\underbrace{\hat{M}_{1}^{(0)},\cdots,\hat{M}_{1}^{(0)}}_{n_1}\Big)
\eeq
where $\hat{M}_{i}^{(0)}=\hat{M}_{i}-\bar{M}^{(0)}$ and 
$\bar{M}^{(0)}=\frac{1}{N}\sum_{j=1}^{N_1}n_j \hat{M}_j$. Considering the parameterization 
(\ref{alpha0}), we have $\beta^{(0)}_i=\hat{M}_i^{(0)}$.

In order to analyze the internal state $\alpha^{(l)}$, we decouple the first $l$ gauge groups 
by setting $c_j=0$ for $j<l$. The Seiberg-Witten curve becomes
\bal
0=&\,\prod_{i=1}^{N}\Big(v-\tilde M_i\Big)z^{k+1}
+c_{k}\Big(v^N-M_k v^{N-1}-u^{(k)}_{2}v^{N-2}-\cdots-u^{(k)}_{N-1}v-u^{(k)}_{N}\Big)z^k +\cdots 
\\ 
&\,+c_l\left(\prod_{i=1}^{N_{l+1}-N_{l}}(v-\hat{M}_i)^{n_i-l}\right) \left(\prod_{i=1}^{N_{l}}(v-r_i)\right) z^l
\eal
where $r_i$, with $i=1,\ldots ,N_l$ are the roots of the polynomial 
\beq
v^{N_l}-M_lv^{N_l-1}-u^{(l)}_{2}v^{N_l-2}-\cdots-u^{(l)}_{N_l-1}v-u^{(l)}_{N_l}
\eeq
and thus satisfy $\sum_{i=1}^{N_l}r_i=M_l$. $r_i$ are the $N_l$ parameters of the Coulomb 
branch of the $SU(N_l)$ gauge group. Defining $\hat{M}_i^{(l)}=\hat{M}_i-\bar{M}^{(l)}$ and 
$r_i^{(l)}=r_i-\bar{M}^{(l)}$ with 
\beq
\bar{M}^{(l)}=\frac{1}{N}\left(\sum_{i=1}^{N_{l+1}-N_{l}}(n_i-l)\hat{M}_i+M_l\right)
\eeq
we can identify semiclassically
\beq
\alpha^{(l)}\sim (r_1^{(l)},\ldots,r_{N_l}^{(l)},
\underbrace{\hat{M}_{N_{l+1}-N_{l}}^{(l)},\ldots, \hat{M}_{N_{l+1}-N_{l}}^{(l)}}_{n_{N_{l+1}-N_{l}}-l},\dots, 
\underbrace{\hat{M}_{1}^{(l)},\ldots,\hat{M}_{1}^{(l)}}_{n_1-l})
\eeq 
The analysis of the Seiberg-Witten curve thus confirm that the internal states $\alpha^{(l)}$ 
are of the form described in (\ref{alphai}).

As mentioned above, the mass parameters as they appear in these states are not the physical 
masses, but are related to them by linear transformations. 
In fact, we can view our identification of states in Section~\ref{sec:subspaces} 
as filling in the information needed to complete the 
writing of the curve in Gaiotto form. The functions $\Phi^{(I)}(z)$ are completely determined by 
knowing their behavior near all their singularities. The leading poles give the W-charges 
of the external states. The lower order poles (like $\phi^{(I)}_{i,-1}$) satisfy certain 
constraints among them \cite{KMST} which mirror the degeneracy conditions of the 
Toda states and after solving for these constraints, the remaining information parametrizes 
the Coulomb branch.

\section{The case of $SU(2)\times SU(3)$}
\label{sec:2x3}

\begin{figure}[!t]
\begin{center}
\vskip4mm
\epsfig{file= 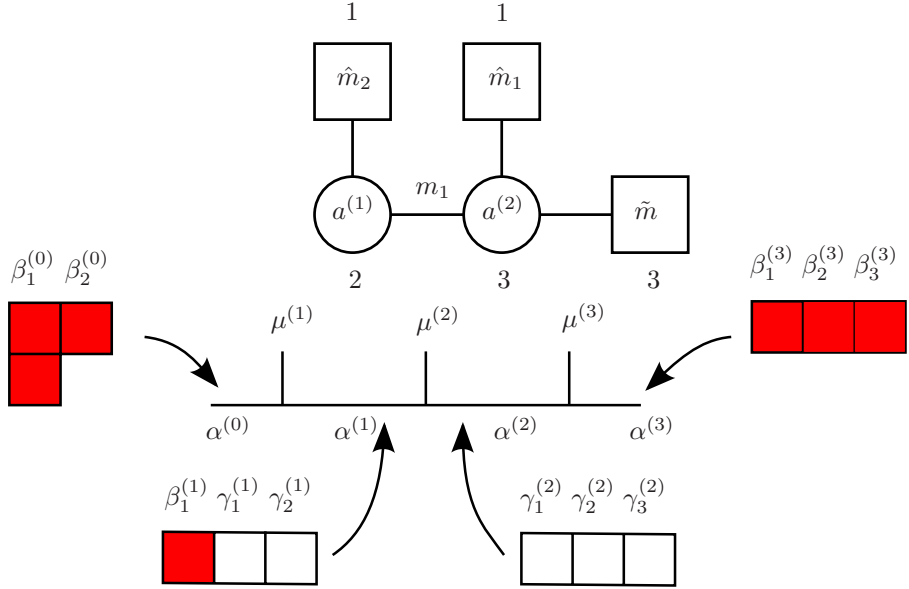,width=120mm
\psfrag{1}{\footnotesize$1$}
\psfrag{2}{\footnotesize$2$}
\psfrag{3}{\footnotesize$3$}
\psfrag{a1}{\footnotesize $a^{(1)}$}
\psfrag{a2}{\footnotesize $a^{(2)}$}
\psfrag{mh1}{\footnotesize $\hat m_1$}
\psfrag{mh2}{\footnotesize $\hat m_2$}
\psfrag{m1}{\footnotesize $m_1$}
\psfrag{a1}{\footnotesize $a^{(1)}$}
\psfrag{a2}{\footnotesize $a^{(2)}$}
\psfrag{a3}{\footnotesize $\tilde m$}
\psfrag{mu1}{\footnotesize $\mu^{(1)}$}
\psfrag{mu2}{\footnotesize $\mu^{(2)}$}
\psfrag{mu3}{\footnotesize $\mu^{(3)}$}
\psfrag{al0}{\footnotesize $\alpha^{(0)}$}
\psfrag{al1}{\footnotesize $\alpha^{(1)}$}
\psfrag{al2}{\footnotesize $\alpha^{(2)}$}
\psfrag{al3}{\footnotesize $\alpha^{(3)}$}
\psfrag{b01}{\footnotesize $\beta^{(0)}_1$}
\psfrag{b02}{\footnotesize $\beta^{(0)}_2$}
\psfrag{b11}{\footnotesize $\beta^{(1)}_1$}
\psfrag{g11}{\footnotesize $\gamma^{(1)}_1$}
\psfrag{g12}{\footnotesize $\gamma^{(1)}_2$}
\psfrag{g21}{\footnotesize $\gamma^{(2)}_1$}
\psfrag{g22}{\footnotesize $\gamma^{(2)}_2$}
\psfrag{g23}{\footnotesize $\gamma^{(2)}_3$}
\psfrag{g31}{\footnotesize $\beta^{(3)}_1$}
\psfrag{g32}{\footnotesize $\beta^{(3)}_2$}
\psfrag{g33}{\footnotesize $\beta^{(3)}_3$}
} 
\parbox{15cm}{
\caption{The simplest linear quiver tail with gauge group $SU(2)\times SU(3)$.
\label{fig:2x3}}}
\end{center}
\end{figure}

We study now in most detail the case of $SU(2)\times SU(3)$ gauge theory, 
whose quiver diagram is shown in Figure~\ref{fig:2x3}. The $SU(2)$ gauge 
group with Coulomb branch parameter $a^{(1)}$ 
couples to a single fundamental field with mass $\hat m_2$ and 
a bi-fundamental field of mass $m_1$. The $SU(3)$ group with Coulomb branch 
parameters $a^{(2)}$ couples to the same 
bi-fundamental field, to a fundamental of mass $\hat m_1$ and three anti-fundamentals 
with masses $\tilde m$.

On the Toda side we have semi-degenerate external states $\alpha^{(0)}$ and 
$\mu^{(1)}$ which fuse to the primary $\alpha^{(1)}$. That is fused with 
$\mu^{(2)}$ to the intermediate state $\alpha^{(2)}$ which is finally fused with 
$\mu^{(3)}$ to the generic external state $\alpha^{(3)}$. We parameterize these 
states as%
\footnote{We omit the subscript from $\beta^{(l)}_1$.}
\beq
\label{2x3states}
\begin{gathered}
\alpha^{(0)}=\vec Q+(-2\beta^{(0)},\beta^{(0)}-Q/2,\beta^{(0)}+Q/2)
\qquad
\vec Q=(Q,0,-Q)
\\
\alpha^{(1)}=\vec Q+(\bar\beta^{(1)}+\gamma^{(1)}_1,\bar\beta^{(1)}+\gamma^{(1)}_2,\beta^{(1)})\,,
\qquad 
\beta^{(1)}=-2\bar\beta^{(1)}\,,
\qquad
\gamma^{(1)}_2=-\gamma^{(1)}_1\,,
\\
\alpha^{(l)}=\vec Q+(\gamma^{(l)}_1,\gamma^{(l)}_2,\gamma^{(l)}_3)\,,
\qquad \sum_{i=1}^3\gamma^{(l)}_i=0\,,
\qquad
l=2,3\,,
\\
\textstyle
\mu^{(l)}=\vec Q+(-2\mu_l,\mu_l-Q/2,\mu_l+Q/2)\qquad l=1,2,3\, .
\end{gathered}
\eeq

\subsection{Fusion rules from Ward identity}
\label{sec:ward}

In the preceding section we proposed a map of the parameters between the quiver tails 
in $\cN=2$ conformal gauge theories and correlation functions in Toda CFT with certain 
semi-degenerate states in the intermediate channels. The space of primary states in the 
intermediate channels is restricted by the conditions \eqn{beta}. Normally we should sum 
over {\em all} primary states in the intermediate channel. Our claim is that only the states 
we wrote down in \eqn{alphai} appear in the fusion rules of the other states (to the left) in 
the correlation function.

In the case of the $SU(2)\times SU(3)$ quiver, this claim amounts to saying that the fusion 
of two semi-degenerate states $\alpha^{(0)}$ and $\mu^{(1)}$ gives a state $\alpha^{(1)}$, 
which while {\em not} degenerate, does satisfy a condition restricting one of its two 
continuous parameters, $\beta^{(1)}$, to be related to $\beta^{(0)}$ and $\mu_1$ 
and in turn, to the masses in the gauge theory picture.

Being semi-degenerate means that the Verma module 
of the state $\alpha^{(0)}$, has a null states at level 1. This degeneracy condition allows 
to write a Ward identity on the three point function of $\alpha^{(0)}$, $\mu^{(1)}$ and 
$\alpha^{(1)}$. Due to the fact that $\mu^{(1)}$ is also semi-degenerate, this Ward 
identity can be written as a condition on the state $\alpha^{(1)}$. 
This Ward identity was written in \cite{KMS} and we present here a slightly different 
derivation of it.

The degeneracy condition of $\alpha^{(0)}$ is given by
\beq\label{deg1}
\langle \alpha^{(0)} |\left(W_1-\frac{3 w_0}{2 \Delta_0} L_1 \right)\sim0
\eeq
where
\beq
\langle \alpha^{(0)} |L_0=\Delta_0\langle \alpha^{(0)} |\,,
\qquad
\langle \alpha^{(0)} |W_0=w_0\langle \alpha^{(0)} |\,.
\eeq

We would like to use the equation satisfied by $\alpha^{(0)}$ to derive a constraint on $\alpha^{(1)}$, 
so we need to commute the operators through $V_{\mu^{(1)}}(z)$.
The action of $L_n$ and $W_n$ on the primaries is given by 
\bal
\label{com}
[L_n,V_{\mu^{(1)}}(z)]&=z^{n+1}\partial V_{\mu^{(1)}} +\Delta_\mu(n+1)z^n V_{\mu^{(1)}}(z)\,,\\
[W_n,V_{\mu^{(1)}}(z)]&=z^{n}\left(\frac{w_\mu}{2}(n+1)(n+2) +(n+2)z\hat{W}_{-1}+z^2 \hat{W}_{-2} \right) V_{\mu^{(1)}}(z)\,.
\eal
By virtue of the degeneracy condition of the state $\mu$, which can be written 
as $[W_{-1},V_{\mu^{(1)}}]=\frac{3w_\mu}{2\Delta_\mu}[L_{-1},V_{\mu^{(1)}}]$, the combinations
\bal
e_n=L_n-zL_{n-1}\,,
\qquad
f_n=W_n-zW_{n-1}-z\frac{3w_\mu}{2\Delta_\mu}L_{n-1}\,,
\eal
have simple commutation relations with $V_{\mu^{(1)}}$
\beq
[e_n,V_{\mu^{(1)}}(z)]=z^n\Delta_\mu V_{\mu^{(1)}}(z)\,,\qquad
[f_n,V_{\mu^{(1)}}(z)]=-z^n\frac{n-2}{2}w_\mu V_{\mu^{(1)}}(z)\,.
\eeq

The degeneracy condition (\ref{deg1}) can be written as 
\beq
\label{degcom}
\langle \alpha^{(0)} |\left(W_1-\frac{3 w_0}{2 \Delta_0} L_1 \right)
=\langle \alpha^{(0)} |\left(f_1-\frac{3 w_0}{2 \Delta_0} e_1
+z\left(-\frac{1}{2}w_0+\frac{3w_\mu}{2\Delta_\mu}\Delta_0\right) \right)\sim0\,.
\eeq
Permuting through $V_{\mu^{(1)}}(z)$ we get
\bal
&\bra{\alpha^{(0)}}\left(W_1-\frac{3 w_0}{2 \Delta_0} L_1 \right)V_{\mu^{(1)}}(z)
\\&\!=\!\bra{\alpha^{(0)}} V_{\mu^{(1)}}\!\!\left(f_1-\frac{3 w_0}{2 \Delta_0} e_1
+z\left(-\frac{1}{2}w_0+\frac{3w_\mu}{2\Delta_\mu}\Delta_0
+\frac{1}{2}w_\mu-\frac{3w_0}{2\Delta_0}\Delta_\mu\right)\right)
\\&\!=\!\bra{\alpha^{(0)}}V_{\mu^{(1)}}\!\!\left(W_1-\frac{3 w_0}{2 \Delta_0} L_1
-z\!\left(W_0+\frac{3}{2}\!\left(\frac{w_\mu}{\Delta_\mu}+\frac{w_0}{\Delta_0}\right)\!L_0
+\frac{w_0}{2}-\frac{3w_\mu}{2\Delta_\mu}\Delta_0
-\frac{w_\mu}{2}+\frac{3w_0}{2\Delta_0}\Delta_\mu\right)\!\right)
\eal
Finally we can write the Ward identity by multiplying by $\ket{\alpha^{(1)}}$ on the right. 
The degeneracy condition implies that this correlation function should vanish. 
Given that $\alpha^{(1)}$ is a primary, it is annihilated by $L_1$ and $W_1$ and 
satisfies $L_0 |\alpha^{(1)}\rangle =\Delta_1 |\alpha^{(1)}\rangle$ and 
$W_0 |\alpha^{(1)}\rangle = w_1 |\alpha^{(1)}\rangle $. We get an equation 
relating $w_1$ and $\Delta_1$ by
\beq
\label{wardprim}
\left(w_1+\frac{3}{2}\left(\frac{w_\mu}{\Delta_\mu}-\frac{w_0}{\Delta_0}\right)(\Delta_1-\Delta_0-\Delta_\mu)
-w_0+w_\mu\right)
\bra{ \alpha^{(0)}} V_{\mu^{(1)}}(z) \ket{\alpha^{(1)}}=0\,.
\eeq
Therefore, for the three point function not to vanish, the prefactor has to. 
This is indeed the equation found by \cite{KMS} which restricts the allowed primaries 
$\alpha^{(1)}$ in the intermediate channel of the $A_2$ description of $SU(2)\times SU(3)$ 
gauge theory.

To understand this equation it is useful to write the quantum numbers of the states as
\bal
\Delta&=\frac{1}{2}\langle\alpha,2\vec Q-\alpha\rangle
=Q^2+\sum_{i<j} \vev{\alpha-Q, h_i}\vev{\alpha-Q, h_j}\,,
\\
w&=\kappa\prod_{i=1}^3\vev{\alpha-Q, h_i}\,,\qquad
\kappa=i\sqrt{\frac{48}{22+5c}}\,.
\eal
Here $h_i$ are the weights of the fundamental representation, which for $SU(3)$ are 
\beq
h_1=\frac{1}{3}(2,-1,-1)\,,\qquad
h_2=\frac{1}{3}(-1,2,-1)\,,\qquad
h_3=\frac{1}{3}(-1,-1,2)\,.
\eeq
For the degenerate states parameterized in the orthonormal basis as in \eqn{2x3states} we 
have
\beq
\begin{gathered}
\Delta_0=3\left(\frac{Q^2}{4}-\big(\beta^{(0)}\big)^2\right)\,,
\qquad
w_0=\frac{2}{3}\kappa\beta^{(0)}\Delta_0\,.
\\
\Delta_\mu=3\left(\frac{Q^2}{4}-\mu_1^2\right)\,,
\qquad
w_\mu=\frac{2}{3}\kappa\mu_1\Delta_\mu\,.
\end{gathered}
\eeq
Plugging these relations into \eqn{wardprim} we find the equation for the components of $\alpha^{(1)}$
\beq
\prod_{i=1}^3\big(\vev{\alpha^{(1)}-Q,h_i}+\mu_1-\beta^{(0)}\big)=0\,.
\eeq
The three different solutions to this equation are related, of course, by Weyl reflections. 
Using the parametrization of $\alpha^{(1)}$ in \eqn{2x3states} we can choose the solution
\beq
\label{beta1}
\beta^{(1)}=\beta^{(0)}-\mu_1\,.
\eeq
As we see, though $\alpha^{(1)}$ is not a degenerate state, one of its components is 
fixed by the external parameters $\beta^{(0)}$ and $\mu_1$.
The other parameter $\gamma^{(1)}_1=-\gamma^{(1)}_2$ is free, and 
should match the Coulomb branch parameter of the $SU(2)$ gauge group.

\subsection{One loop contribution}
\label{sec:2x31-loop}

Within the AGT correspondence the product of the 1-loop determinants of the vector and hypermultiplets 
matches the three point functions associated to the fusion of successive primary states in the CFT. 

In the following two sections we study the three point functions of semi-degenerate states in 
Toda CFT which are needed for the calculation of the quiver tail. In particular we show in 
Section~\ref{sec:3pt} that 
imposing the condition \eqn{beta1} gives a pole in the three point function, whose residue is 
the desired three point function. In Section~\ref{sec:together} we combine all the terms 
together and show in a more general setup that the 1-loop partition function agrees with the 
product of three point functions in Toda CFT. It is easy to plug in the states parameterized 
above in \eqn{2x3states} into the resulting expressions, and we will not copy it here.

The upshot of that calculation is that it allows us to identify the gauge theory and Toda 
parameters as
\bal\label{dictio}
&\gamma^{(1)}_i=a^{(1)}_i\,,
&\qquad
&\gamma^{(2)}_i=a^{(2)}_i\,,
&\qquad
&\beta^{(3)}_i+\mu_3=\tilde m_i\,,
\\
&3\beta^{(0)}=m_1+\hat m_1-\hat m_2\,,
&\qquad
&\frac{3}{2}\beta^{(1)}=m_1+\hat m_1\,,
\\
&3\mu_1=-m_1-\hat m_1-\hat m_2\,,
&\qquad
&3\mu_2=2m_1-\hat m_1\,,
&\qquad
&3\mu_3=\sum_{i=1}^3\tilde m_i\,.
\eal

\subsection{Instantons and descendants}
\label{sec:2x3inst}

Within the AGT correspondence, the instanton corrections to the gauge theory partition 
function \cite{Nekrasov:2002qd, Nekrasov:2003rj} match the contributions of descendants to the conformal blocks of the CFT. We 
turn now to analyzing this question for our case of the $SU(2)\times SU(3)$ gauge theory 
and its description within $A_2$ Toda CFT.

The relevant Toda correlator we are calculating is on the five-punctured sphere represented 
by the graphs in Figure~\ref{fig:2x3}. There are two intermediate states, $\alpha^{(1)}$ and 
$\alpha^{(2)}$ and the conformal blocks capture the contributions of all W-algebra descendants 
of these primaries
\bal
\cF(q^{(1)},q^{(2)})=\sum_{{\bf x},{\bf x}',{\bf y},{\bf y}'} (q^{(1)})^{|{\bf x}|} (q^{(2)})^{|{\bf y}|}
&
\frac{\langle \alpha^{(0)} | V_{\mu^{(1)}} |{\bf x}; \alpha^{(1)} \rangle}
{\langle \alpha^{(0)} | V_{\mu^{(1)}} | \alpha^{(1)} \rangle} X^{-1}_{ \bf x; \bf x'}(\alpha^{(1)})
\frac{\langle {\bf x}'; \alpha^{(1)} |V_{\mu^{(2)}} |{\bf y};\alpha^{(2)} \rangle}
{\langle \alpha^{(1)} |V_{\mu^{(2)}} |\alpha^{(2)} \rangle} 
\\&\times
X^{-1}_{ \bf y; \bf y'}(\alpha^{(2)})
\frac{\langle {\bf y}'; \alpha^{(2)} |V_{\mu^{(3)}} |\alpha^{(3)}\rangle}
{\langle \alpha^{(2)} |V_{\mu^{(3)}} |\alpha^{(3)}\rangle} \,.
\eal
The expansion parameters $q^{(i)}$ are the complex structure moduli of the Riemann 
surface, related to the gauge groups couplings $\tau^{(i)}$ by $q^{(i)}=e^{2\pi i \tau^{(i)}}$. 
The sum is taken over all the descendants 
$ |{\bf x};\alpha\rangle=L_{-x_{1}}\ldots L_{-x_{r}}W_{-\tilde{x}_{1}}\ldots W_{-\tilde{x}_{\tilde{r}}}|\alpha\rangle$, represented by the vectors ${\bf x}=(x_1,\ldots, x_r, \tilde{x}_1,\ldots, \tilde{x}_{\tilde{r}})$ with 
$x_1\leq\dots\leq x_{r}$ and $\tilde{x}_1\leq\dots\leq \tilde{x}_{r}$. 
The level of the descendant is given by 
$|{\bf x}|=x_1+\ldots+x_r+\tilde{x}_1+\ldots+\tilde{x}_{\tilde{r}}$. 
$X^{-1}_{ \bf x; \bf x'}(\alpha)$ is the inverse of the Shapovalov matrix defined as 
$X_{ \bf x; \bf x'}(\alpha)= \langle \bf x; \alpha |\bf x' ; \alpha \rangle $. The Shapovalov 
matrix is block diagonal, where any block correspond to a set of descendants with a given level. 
Restricting to a fixed level we define
\bal
\cF(q^{(1)},q^{(2)})
=\sum_{n_1,n_2}\cF_{n_1,n_2}(q^{(1)})^{n_1} (q^{(2)})^{n_2}\, .
\eal

The contribution of a single instanton to the $SU(3)$ gauge theory was matched with the level 
one conformal block in $A_2$ Toda CFT in \cite{Wyllard:2009hg} and two instantons were matched 
with level two in \cite{Mironov:2009by}. The generalization to the 
linear quiver with $SU(3)\times SU(3)$ gauge group was done in \cite{KMS}. Here we 
study the case of $SU(2)\times SU(3)$ and the novel question, as with the primaries 
discussed above, is how does $A_2$ Toda reproduce the $SU(2)$ part of the theory, which 
is also captured by Liouville theory. To address this question we can focus on the 
$q^{(1)}$ dependance and ignore all descendants of $\alpha^{(2)}$
\beq
\label{wblock1}
{\cal F}(q^{(1)})=\sum_{n_1}\cF_{n_1,0}(q^{(1)})^{n_1}
=\sum_{{\bf x},{\bf x}'} (q^{(1)})^{|{\bf x}|} \frac{\langle \alpha^{(0)} | V_{\mu^{(1)}} |{\bf x}; \alpha^{(1)} \rangle}{\langle \alpha^{(0)} | V_{\mu^{(1)}} | \alpha^{(1)} \rangle} X^{-1}_{ \bf x; \bf x'}(\alpha^{(1)})
\frac{\langle {\bf x}'; \alpha^{(1)} |V_{\mu^{(2)}} |\alpha^{(2)} \rangle}{\langle \alpha^{(1)} |V_{\mu^{(2)}} |\alpha^{(2)} \rangle}\,.
\eeq 

We will restrict our analysis to level one, where there are two descendants 
$L_{-1}|\alpha^{(1)}\rangle$ and $W_{-1}|\alpha^{(1)}\rangle$. The Shapovalov matrix for the level one states 
is a $2\times2$ matrix which was evaluated in \cite{Wyllard:2009hg, Mironov:2009by}, as were the ratio of 3-point 
functions of the descendants and primaries. We plug in the expression for the state 
$\alpha^{(1)}$ satisfying the condition \eqn{beta1} and use the degeneracy condition for 
$\alpha^{(0)}$ and $\mu^{(1)}$ and $\mu^{(2)}$ to find 
(in the parametrization \eqn{2x3states})
\beq
\label{F10}
\cF_{1,0}=-6\frac{(\beta^{(0)}-\frac{\beta^{(1)}}{2})
\prod_{i=1}^3(\gamma^{(2)}_i+\mu_2+\frac{\beta^{(1)}}{2})}
{Q^2-4(\gamma^{(1)}_1)^2}
+D\,,
\eeq
where
\beq
D=\frac{1}{4}\sum_{i=1}^3(\gamma^{(2)}_i)^2-\frac{1}{2}(\gamma^{(1)}_1)^2
+\frac{3}{4}((\beta^{(0)})^2-2\mu_2^2-2\mu_1^2+\beta^{(0)}\mu_1+\beta^{(0)}\mu_2+\mu_1\mu_2)+\frac{3}{8}Q^2
\eeq

In the gauge theory the $SU(2)$ vector multiplet with Coulomb branch parameter $a^{(1)}$ 
couples to one fundamental and three bi-fundamental fields. 
This is equivalent to four fundamental hypermultiples 
with masses equal to $\hat{m}_2$, $m_1+a_1^{(2)}$, $m_1+a_2^{(2)}$, $m_1+a_3^{(2)}$. 
An explicit instanton counting for $SU(2)$ gauge theory coupled to 4 fundamental 
hypermultiplet with these masses gives \cite{Nekrasov:2004vw, Marino:2004cn,Wyllard:2009hg}
\beq
\label{ins}
Z_\text{inst}(q^{(1)})=1+q^{(1)}\frac{2\hat m_2\prod_{i=1}^3(a^{(2)}_i+m^{(2)})}{Q^2-4a^{(1)}_1}+{\cal O}((q^{(1)})^2)
\eeq
Using the dictionary relating the Toda parameters and the gauge theory parameters (\ref{dictio}), 
derived from comparing the one-loop partition function and Toda three point functions, 
we find that the ratio in \eqn{F10} matches exactly the partition function for a single instanton in 
$SU(2)$ \eqn{ins}. It is tempting to conjecture that a similar relation persists to higher order 
in the instanton expansion such that the $SU(2)$ instanton partition functions and 
$A_2$ Toda conformal blocks would be related by
\beq
{\cal F}(q^{(1)})=(1-q^{(1)})^{-D}Z_\text{inst}(q^{(1)})
\eeq 
This is essentially the same%
\footnote{With a different $D$.}
as we would get by considering Liouville theory,
where at level one there is a single state in the Virasoro Verma module. 
The relation we derived involved summing over {\em two} W-algebra descendants and, 
unlike in Liouville, the denominator in \eqn{F10} is {\em not} the conformal dimension 
of the Toda state $\alpha^{(1)}$, yet it agrees with the Liouville dimension and gauge 
theory expression.

\subsection{The Seiberg-Witten curve for $SU(2)\times SU(3)$}
\label{sec:SW2x3}

The Seiberg-Witten curve \eqn{witt} for the $SU(2)\times SU(3)$ theory is 
\bal
\label{witt2x3}
0=&\,\big(v-\tilde M_1\big)\big(v-\tilde M_2\big)\big(v-\tilde M_3\big)z^{3}
+c_{2}\big(v^3-M_2 v^2-u^{(2)}_{2}v-u^{(2)}_{3}\big)z^2
\\&\,+c_1\big(v-\hat{M}_1)(v^{2}-M_1v-u^{(1)}_{2}\big)z
+c_0(v-\hat{M}_1)^{2}(v-\hat{M}_2)\,.
\eal
We can easily follow the procedure outlined in Section~\ref{sec:SW} and write this 
curve in Gaiotto form as
\beq
\label{gaio2x3}
x^3 +\Phi^{(2)}(z)x+\Phi^{(3)}(z)=0 
\eeq
$\Phi^{(I)}(z)$ has poles at $z=0,1,A,B,\infty$ where we set
\beq
c_0=-AB\,,
\qquad
c_1=A+B+AB\,,
\qquad
c_2=-1-A-B\,.
\eeq
The parameters of the Toda states are related to the Laurent coefficients 
of $\Phi^{(I)}$ giving
\bal
\mu_1
&=\frac{(B-A-AB)\hat{M}_1+B\hat{M}_2-(A+B+AB)M_1+A(1+A+B)M_2-A^2\sum_{i=1}^3\tilde M_i}{3(1-A)(A-B)}\,,
\\
\mu_2&=\frac{(A+B-AB)\hat{M}_1-AB\hat{M}_2+(A+B+AB)M_1-(1+A+B)M_2+\sum_{i=1}^3\tilde M_i}{3(1-A)(1-B)}\,,
\\
\mu_3
&=\frac{(A-B-AB)\hat{M}_1+A\hat{M}_2-(A+B+AB)M_1+B(1+A+B)M_2-B^2\sum_{i=1}^3\tilde M_i}{3(A-B)(1-B)}\,,
\eal
As mentioned in Section~\ref{sec:SW} (see footnote~\ref{massnote}),  these complicated 
relations are a consequence of a complicated relation between the physical masses and the 
parameters in \eqn{witt2x3} (as well as the shift in $v$ to \eqn{gaio2x3}). These expressions 
agree with our identifications in \eqn{dictio} once we relate the parameters in \eqn{witt2x3} 
to the physical masses as
\bal
\label{mred}
M_1&=
\frac{A\hat m_1+A(1+B)\hat m_2-(A+2B-AB)m_1+\frac{1}{3}(A+4B+4AB)\sum_{i=1}^{3}\tilde m_i}{A+B+AB}\,,
\\
M_2&=
\frac{(1+A)\hat{m}_1A\hat{m}_2+(A-2)m_1+(1+A+2B)\sum_{i=1}^{3}\tilde m_i}{1+A+B}\,,
\\
\hat{M}_1&=
\hat{m}_1+\frac{2}{3}\sum_{i=1}^{3}\tilde m_i\,,
\qquad
\hat{M}_2=
\hat{m}_2-m_1+\frac{2}{3}\sum_{i=1}^{3}\tilde m_i\,,
\qquad
\tilde M_i=\tilde m_i\,.
\eal
With this, the parameters $\mu_i$ as well as $\beta^{(0)}$ satisfy the relations 
in \eqn{dictio}.

According to our prescription, of the two internal states in the Toda description of this 
theory $\alpha^{(1)}$ has one parameter $\beta^{(1)}$ fixed by the external data and one 
free parameter and $\alpha^{(2)}$ is completely free. In Section~\ref{sec:SW} we 
explained how to see the constraint on the internal states from the Seiberg-Witten 
curve. We consider the limit of $c_0=0$ in \eqn{witt2x3} which gives
\beq
0=\prod_{i=1}^{3}\Big(v-\tilde M_i\Big)z^{3}+c_{2}\Big(v^3-u^{(2)}_{2}v-u^{(2)}_{3}\Big)z^2
+c_1(v-\hat{M}_1)(v-r_1)(v-r_2)z
\eeq
where $r_1$, $r_2$ are the roots of the polynomial 
\beq
v^{2}-M_1v-u^{(1)}_{2}
\eeq
and thus satisfy $r_1+r_2=M_1$. Defining $\hat{M}_1^{(1)}=\hat{M}_1-\bar{M}^{(1)}$ and $r_i^{(1)}=r_i-\bar{M}^{(1)}$ with $\bar{M}^{(1)}=\frac{1}{3}(\hat{M}_1+M_1)$ we can identify semiclassically
\beq
\alpha^{(1)}\sim (r_1^{(1)},r_{2}^{(1)},\hat{M}_{1}^{(1)})
\eeq 
In terms of our parametrization of the states in \eqn{2x3states} we have
\beq
\label{bo}
\beta^{(1)}=\hat{M}^{(1)}_1
=\frac{1}{3}\left(2\hat{M}_1-M_1\right)
\eeq
After performing the redefinition \eqn{mred} we get
\beq
\label{bb} 
\beta^{(1)}=\frac{2}{3}(\hat m_1+m_1)\,,
\eeq
as in equation \eqn{dictio}.

\section{Three point functions of semi-degenerate states}
\label{sec:3pt}

In the previous section we showed how a Ward identity constrains the fusion of a 
semi-degenerate state and a simple puncture state to a third state satisfying certain 
restrictions on its parameters. The derivation involved details of the $W_3$ algebra. 
In this section we show a simpler route to the same conclusion. We look at the 
three point functions of two generic states and one simple puncture conjectured by 
Fateev and Litvinov \cite{Fateev:2005gs, Fateev:2007ab} 
and consider the limit when the two generic states become semi-degenerate.

While one would expect to find a pole in the three point function, as we shall see, 
generically the result is zero. To cancel this zero we impose the condition 
(\ref{beta}) on the momenta of the states and find extra zeroes in the denominator, 
giving the expected pole, whose coefficient is then interpreted as the three point 
function of the semi-degenerate states.

The three point function of two generic and one simple puncture state 
proportional to the last fundamental weight 
is \cite{Fateev:2005gs, Fateev:2007ab}
\bal
\label{fl-anti}
C_{FL}\big(\alpha,({\textstyle\frac{Q}{2}}-\mu)&N\omega_{N-1},\alpha'\big)=
\left[\pi\bar\mu\gamma(b^2)b^{2-2b^2}\right]^{\langle2\vec Q-\alpha-\alpha',\rho\rangle/b}
\\ &
\times \frac{\left(\Upsilon(b)\right)^{N-1}
\Upsilon\big(N({\textstyle\frac{Q}{2}}-\mu)\big)\prod_{e>0}\Upsilon\big(\langle \vec Q-\alpha,e\rangle\big)
\Upsilon\big(\langle \vec Q-\alpha',e\rangle\big)}
{\prod_{ij}\Upsilon\big(\frac{Q}{2}-\mu+
\langle\alpha-\vec Q,h_i\rangle+\langle\alpha'-\vec Q,h_j \rangle\big)}\,.
\eal
The product in the numerator is over all the positive roots and in the denominator over 
all the weights of the fundamental representation. The two special functions in this 
expression are $\gamma(x)=\Gamma(x)/\Gamma(1-x)$ and $\Upsilon(x)$, which is 
defined in the appendix. $\bar\mu$ is the cosmological constant of Toda theory, not 
to be confused with the parameter of the state $\frac{Q}{2}-\mu$.

For the 
simple puncture proportional to the first fundamental weight 
the expression is
\bal
\label{fl}
C_{FL}\big(\alpha,{\textstyle(\frac{Q}{2}-\mu)}&N\omega_1,\alpha'\big)=
\left[\pi\bar\mu\gamma(b^2)b^{2-2b^2}\right]^{\langle2\vec Q-\alpha-\alpha',\rho\rangle/b}
\\ &
\times \frac{\left(\Upsilon(b)\right)^{N-1}
\Upsilon\big(N({\textstyle\frac{Q}{2}}-\mu)\big)\prod_{e>0}\Upsilon\big(\langle \vec Q-\alpha,e\rangle\big)
\Upsilon\big(\langle \vec Q-\alpha',e\rangle\big)}
{\prod_{ij}\Upsilon\big(\frac{Q}{2}-\mu-
\langle\alpha-\vec Q,h_i\rangle-\langle\alpha'-\vec Q,h_j \rangle\big)}\,.
\eal
Using the reflection relation $\Upsilon(x)=\Upsilon(Q-x)$, we see that the denominators 
of the above expressions are equal upon replacement of $\mu\to-\mu$.

Now we take $\alpha$ and $\alpha'$ to be semi-degenerate states. 
Since keeping track of indices is really tedious in this calculation we will concentrate 
on the case when the two states are ``hooks'', {\em i.e.,} the Young-diagrams have only 
one column of height greater than one. We consider the state $\alpha$ with Young-diagram 
$[n, 1,\cdots, 1]$ and $\alpha'$ with Young-diagram $[n',1,\cdots,1]$%
\footnote{In comparison with our notation in \eqn{alphai}, we 
omit the subscript from $\beta_1$ and take $\bar\beta=-n\beta/(N-n)$.}
\beq
\label{hook}
\textstyle
\alpha=\vec Q+\big(-\frac{n}{N-n}\beta+\gamma_1,\cdots ,
-\frac{n}{N-n}\beta+\gamma_{N-n},\beta+\delta_{n,1},\cdots,\beta+\delta_{n,n}\big)
\eeq
and likewise $\alpha'$.

Specializing to the ``hook'' states $\alpha$ and $\alpha'$ we find
\bal
\label{hookncb}
C_{FL}\big(2\vec Q-\alpha,{\textstyle(\frac{Q}{2}-\mu)}&N\omega_1,\alpha'\big)=
\left[\pi\bar\mu\gamma(b^2)b^{2-2b^2}\right]^{\langle \alpha-\alpha',\rho \rangle/b}
\left(\Upsilon(b)\right)^{N-1}
\Upsilon\big(N({\textstyle\frac{Q}{2}}-\mu)\big)
\\&\times
\frac{\prod_{i<j\leq N-n}\Upsilon\big(\gamma_i-\gamma_j\big)
\prod_{i<j\leq N-n'}\Upsilon\big(\gamma_j'-\gamma_i'\big)}
{\prod_{i\leq N-n}\prod_{j\leq N-n'}\Upsilon\big(\frac{Q}{2}-\mu-\frac{n}{N-n}\beta
+\frac{n'}{N-n'}\beta'+\gamma_i-\gamma_j'\big)}
\\&\times
\frac{\prod_{j \leq n}\prod_{i \leq N-n}\Upsilon\big(-\frac{N }{N-n}\beta+\gamma_i-\delta_{n,j}\big)}
{\prod_{i\leq N-n}\prod_{j \leq n'}\Upsilon\big(\frac{Q}{2}-\mu-\frac{n}{N-n}\beta
-\beta'+\gamma_i-\delta_{n',j}\big)}
\\ &\times
\frac{\prod_{i \leq N-n'}\prod_{j \leq n'}\Upsilon\big(\frac{N}{N-n'}\beta'-\gamma_i'+\delta_{n',j}\big)}
{\prod_{j\leq N-n'}\prod_{i \leq n}\Upsilon\big(\frac{Q}{2}-\mu+\beta
+\frac{n'}{N-n'}\beta'-\gamma_j'+\delta_{n,i}\big)}
\\&\times
\frac{\prod_{i < j\leq n}\Upsilon(\delta_{n,i}-\delta_{n,j})\prod_{i < j\leq n'}\Upsilon(\delta_{n',j}-\delta_{n',i})}
{\prod_{i\leq n}\prod_{j \leq n'}\Upsilon\big(\frac{Q}{2}-\mu+\beta-\beta'
+\delta_{n,i}-\delta_{n',j}\big)}
\eal

Recall that $\delta_{n,j}=(j-(n+1)/2)Q$, such that the difference $\delta_{n,j}-\delta_{n,i}$ is an 
integer multiple of $Q$, where $\Upsilon$ has a zero. Therefore the numerator on the last line 
of \eqn{hookncb} has a zero of order $\frac{1}{2}(n(n-1)+n'(n'-1))$.

We found that for generic values of the momenta $\beta$ and $\gamma$ the 3-point function vanishes. 
For specific values of the momentum we can get extra zeroes in the denominator which may 
cancel the numerator. If we impose%
\footnote{Note that in our notations, $\beta$, $\beta'$ and $\mu$ are all purely imaginary.}
\beq
\label{constraint}
\beta-\beta'=\mu\,,
\eeq
and if $n-n'$ is odd, the argument of the Upsilon functions in the denominator of the last line 
of \eqn{hookncb} will also be integer multiples of $Q$ giving $nn'$ zeros in the denominator and 
the overall order of the zero is
\beq
\label{zero}
\frac{(n-n')^2-n-n'}{2}\,.
\eeq
Let us assume $n>n'$, then equation \eqn{constraint} represents $n'$ conditions on the state 
$\alpha'$. This should correspond to a pole of order $n'$ in the three point function. 
Solving that \eqn{zero} is $-n'$ yields
\beq
n'=n-1\,.
\eeq

If we consider two general semi-degenerate states a similar argument can be applied 
separately to each column. This confirms our claim in Section~\ref{sec:tails} that 
the consecutive states in the intermediate channels $\alpha^{(l)}$ correspond to 
Young-diagrams with all the columns of length greater than one, shortened by one, 
and the first row extended accordingly.

Going back to the ``hook'' states with $n'=n-1$, using \eqn{constraint}, the three point function 
\eqn{hookncb} can now be simplified as
\bal
C_{FL}&\big(2\vec Q-\alpha,{\textstyle(\frac{Q}{2}-\mu)}N\omega_{1},\alpha'\big)=
\left[\pi\bar\mu\gamma(b^2)b^{2-2b^2}\right]^{\langle \alpha-\alpha',\rho \rangle/b}
\left(\Upsilon(b)\right)^{N-1}
\Upsilon\big(N({\textstyle\frac{Q}{2}}-\mu)\big)
\\&\times
\frac{\Upsilon(nQ)}{\Upsilon(0)^n}
\frac{\prod_{i<j\leq N-n}\Upsilon\big(\gamma_i-\gamma_j\big)
\prod_{i<j\leq N-n'}\Upsilon\big(\gamma_j'-\gamma_i'\big)}
{\prod_{i\leq N-n}\prod_{j\leq N-n'}\Upsilon\big(\frac{Q}{2}-\frac{N}{N-n}\beta
+\frac{N}{N-n'}\beta'+\gamma_i-\gamma_j'\big)}
\\&\times
\frac{\prod_{i \leq N-n}\prod_{j \leq n}\Upsilon\big(-\frac{N }{N-n}\beta+\gamma_i-\delta_{n,j}\big)}
{\prod_{i\leq N-n}\prod_{j \leq n'}\Upsilon\big(\frac{Q}{2}-\frac{N }{N-n}\beta+\gamma_i-\delta_{n',j}\big)}
\frac{\prod_{i \leq N-n'}\prod_{j \leq n'}\Upsilon\big(\frac{N}{N-n'}\beta'-\gamma_i'+\delta_{n',j}\big)}
{\prod_{i\leq N-n'}\prod_{j \leq n}\Upsilon\big(\frac{Q}{2}+\frac{N}{N-n'}\beta'-\gamma_i'+\delta_{n,j}\big)}
\eal
Using that $\Upsilon(Q-x)=\Upsilon(x)$ this can be further simplified to
\bal
\label{3pt}
C_{FL}\big(2\vec Q-\alpha,{\textstyle(\frac{Q}{2}-\mu)}&N\omega_{1},\alpha'\big)=
\left[\pi\bar\mu\gamma(b^2)b^{2-2b^2}\right]^{\langle \alpha-\alpha',\rho \rangle/b}
\left(\Upsilon(b)\right)^{N-1}
\Upsilon\big(N({\textstyle\frac{Q}{2}}-\mu)\big)
\\&\times
\frac{\Upsilon(nQ)}{\Upsilon(0)^n}
\frac{\prod_{i<j\leq N-n}\Upsilon\big(\gamma_i-\gamma_j\big)
\prod_{i<j\leq N-n'}\Upsilon\big(\gamma_j'-\gamma_i'\big)}
{\prod_{i\leq N-n}\prod_{j\leq N-n'}\Upsilon\big(\frac{Q}{2}+\frac{N}{N-n}\beta
-\frac{N}{N-n'}\beta'-\gamma_i+\gamma_j'\big)}
\\&\times
\frac{\prod_{i \leq N-n}\Upsilon\big(\frac{N}{N-n}\beta-\gamma_i+\frac{n+1}{2}Q\big)}
{\prod_{i\leq N-n'}\Upsilon \big(\frac{N}{N-n'}\beta'-\gamma_i'+\frac{n'+1}{2}Q\big)}
\eal
As explained above, this expression has a pole of order $n-1$, and the desired three point 
function is the residue at the pole.

\section{Quiver tail in Toda CFT}
\label{sec:together}

We are now ready to check the one-loop part of the 
AGT correspondence, and our identification of the 
space of allowed intermediate primary states. We employ the formula derived in 
the previous section and therefore consider a ``simple tail''. That is 
our name for the gauge theory with $SU(2)\times SU(3)\times\cdots\times SU(N)$ 
gauge symmetry, such that the Riemann surface has one full puncture and $N+1$ 
simple punctures.

\subsection{1-loop partition function}
\label{sec:1loop}

We start with the gauge theory calculation. There are $N-1$ vector multiplets 
with Coulomb branch parameters $a^{(l)}$, $l=1,\cdots N-1$. Together with 
the Vandermonde determinant, each vector multiplet contributes to the 1-loop partition 
function \cite{Nekrasov:2002qd, Nekrasov:2003rj}
\beq
\label{vector}
|Z^\text{vector}_\text{1-loop}|^2=
\frac{\prod_{i<j}\big|a^{(l)}_i-a^{(l)}_j\big|^2}
{\prod_{i<j}\big|\Gamma_b\big(a^{(l)}_i-a^{(l)}_j+1/b\big)\Gamma_b\big(a^{(l)}_i-a^{(l)}_j+b\big)\big|^2}
=\prod_{i<j}\Upsilon\big(a^{(l)}_i-a^{(l)}_j\big)\Upsilon\big(a^{(l)}_j-a^{(l)}_i\big)\,.
\eeq
The bi-fundamental fields charged under $SU(l)\times SU(l+1)$ contribute
\beq
\label{bi-fund}
|Z^\text{bi-fund}_\text{1-loop}|^2
=\prod_{i=1}^{l}\prod_{j=1}^{l+1}
\big|\Gamma_b\big({\textstyle\frac{Q}{2}}+a^{(l-1)}_i-a^{(l)}_j-m_{l-1}\big)\big|^2
=\frac{1}{\prod_{i=1}^{l}\prod_{j=1}^{l+1}\Upsilon\big({\textstyle\frac{Q}{2}}+a^{(l-1)}_i-a^{(l)}_j-m_{l-1}\big)}
\eeq

Lastly we have the contributions of the fundamental and anti-fundamental fields. For the 
``simple quiver tail'' there is one fundamental field charged under $SU(2)$, one under 
$SU(N)$ and $N$ anti-fundamental fields also charged under $SU(N)$. Their contribution 
is
\bal
\label{fund}
|Z^\text{fund}_\text{1-loop}|^2
&=\prod_{i=1}^2\big|\Gamma_b\big({\textstyle\frac{Q}{2}}+a^{(1)}_i-\hat m_2\big)\big|^2
\prod_{i=1}^N
\big|\Gamma_b\big({\textstyle\frac{Q}{2}}+a^{(N-1)}_i-\hat m_1\big)\big|^2
\\&\qquad\times
\prod_{i=1}^N\prod_{j=1}^{N}
\big|\Gamma_b\big({\textstyle\frac{Q}{2}}-a^{(N-1)}_i+\tilde m_j\big)\big|^2
\\&
=\frac{1}{
\prod_{i=1}^2\Upsilon\big({\textstyle\frac{Q}{2}}+a^{(1)}_i-\hat m_2\big)
\prod_{i=1}^N
\Upsilon\big({\textstyle\frac{Q}{2}}+a^{(N-1)}_i-\hat m_1\big)}
\\&\qquad\times
\frac{1}{\prod_{i=1}^N\prod_{j=1}^{N}
\Upsilon\big({\textstyle\frac{Q}{2}}-a^{(N-1)}_i+\tilde m_j\big)}
\eal

The full 1-loop contribution is the product of the $N-1$ vector multiplets \eqn{vector}, 
$N-2$ bi-fundamental fields \eqn{bi-fund} and the fundamentals \eqn{fund}. We now 
show how these terms arise from the 3-point functions of semi-degenerate states in 
Toda CFT.

\subsection{Product of Toda 3-point functions}
\label{sec:toda-tail}

According to our prescription the external states are $\alpha^{(0)}$, $\alpha^{(N)}$ and 
$\mu^{(l)}$ and the states in the intermediate channels are $\alpha^{(l)}$. The states 
are parameterized as in \eqn{alphai}, which is the same as \eqn{hook} with 
$n_l=N-l-1$. The states $\mu^{(l)}=(Q/2-\mu_l)N\omega_1$ and the intermediate 
states satisfy the constraints \eqn{constraint} . We use \eqn{3pt} to evaluate the 
consecutive three point functions. Up to a momentum independent constant 
the first three point function is
\bal
C_{FL}(2\vec Q-\alpha^{(0)},&\mu^{(1)},\alpha^{(1)})\propto
\left[\pi\bar\mu\gamma(b^2)b^{2-2b^2}\right]^{\langle \alpha^{(0)}-\alpha^{(1)},\rho \rangle/b}
\Upsilon\big(N({\textstyle\frac{Q}{2}}-\mu_1)\big)
\\&\times
\frac{\Upsilon\big(N(\frac{Q}{2}+\beta^{(0)})\big)\prod_{i<j\leq 2}\Upsilon\big(\gamma^{(1)}_j-\gamma^{(1)}_i\big)}
{\prod_{j\leq 2}\Upsilon\big(\frac{Q}{2}+N\beta^{(0)}
-\frac{N}{2}\beta^{(1)}+\gamma^{(1)}_j\big)
\prod_{i\leq2}\Upsilon \big(\frac{N}{2}\beta^{(1)}-\gamma^{(1)}_i+\frac{N-1}{2}Q\big)}
\eal
The next $N-3$ three point functions, with $l=2,\cdots N-2$, are of the form
\bal
\label{3pt-int}
C_{FL}(2\vec Q-\alpha^{(l-1)},\mu^{(l)},\alpha^{(l)})\propto&\,
\left[\pi\bar\mu\gamma(b^2)b^{2-2b^2}\right]^{\langle \alpha^{(l-1)}-\alpha^{(l)},\rho \rangle/b}
\Upsilon\big(N({\textstyle\frac{Q}{2}}-\mu_l)\big)
\\&
\times\frac{\prod_{i<j\leq l}\Upsilon\big(\gamma^{(l-1)}_i-\gamma^{(l-1)}_j\big)
\prod_{i<j\leq l+1}\Upsilon\big(\gamma^{(l)}_j-\gamma^{(l)}_i\big)}
{\prod_{i\leq l}\prod_{j\leq l+1}\Upsilon\big(\frac{Q}{2}+\frac{N}{l}\beta^{(l-1)}
-\frac{N}{l+1}\beta^{(l)}-\gamma^{(l-1)}_i+\gamma_j^{(l)}\big)}
\\&\times
\frac{\prod_{i \leq l}\Upsilon\big(\frac{N }{l}\beta^{(l-1)}-\gamma^{(l-1)}_i+\frac{N-l+1}{2}Q\big)}
{\prod_{i\leq l+1}\Upsilon \big(\frac{N}{l+1}\beta^{(l)}-\gamma^{(l)}_i+\frac{N-l}{2}Q\big)}
\eal
Note that the ratio on the last line will cancel between consecutive terms 
in the product over $l$ and likewise the prefactor in the square bracket. 
The last two intermediate states, $\alpha^{(N-2)}$ and 
$\alpha^{(N-1)}$ are not degenerate. $\alpha^{(N-2)}$ still has one component $\beta^{(N-2)}$ 
and $N-1$ $\gamma^{(N-2)}_i$ components. For $\alpha^{(N-1)}$ there is no $\beta^{(N-1)}$, 
but only $\gamma^{(N-1)}_i$. The three point function is still the same as above, with $l=N-1$ if we 
define $\beta^{(N-1)}=\beta^{(N-2)}-\mu_{N-1}$.

The last three point function is between two non-degenerate states and a simple puncture
\bal
C_{FL}(2\vec Q-\alpha^{(N-1)},&\mu^{(N)},\alpha^{(N)})\propto
\left[\pi\bar\mu\gamma(b^2)b^{2-2b^2}\right]^{\langle \alpha^{(N-1)}-\alpha^{(N)},\rho \rangle/b}
\\&\times
\frac{\Upsilon\big(N({\textstyle\frac{Q}{2}}-\mu_N)\big)
\prod_{i<j\leq N}\Upsilon\big(\gamma^{(N-1)}_i-\gamma^{(N-1)}_j\big)
\prod_{i<j\leq N}\Upsilon\big(\beta^{(N)}_j-\beta^{(N)}_i\big)}
{\prod_{i\leq N}\prod_{j\leq N}\Upsilon\big(\frac{Q}{2}-\mu_N+\gamma^{(N-1)}_i-\beta^{(N)}_j\big)}
\eal

Combining all the terms together we find (with $\gamma^{(0)}=0$)
\bal
\prod_{l=1}^N&C_{FL}(2\vec Q-\alpha^{(l-1)},\mu^{(l)},\alpha^{(l)})
\propto
\left[\pi\bar\mu\gamma(b^2)b^{2-2b^2}\right]^{\langle \alpha^{(0)}-\alpha^{(N)},\rho \rangle/b}
\\&\times
\prod_{l=1}^{N-1}\frac{
\prod_{i<j\leq l}\big|\Upsilon\big(\gamma^{(l)}_i-\gamma^{(l)}_j\big)\big|^2}
{\prod_{i\leq l}\prod_{j\leq l+1}\Upsilon\big(\frac{Q}{2}+\frac{N}{l}\beta^{(l-1)}
-\frac{N}{l+1}\beta^{(l)}-\gamma^{(l-1)}_i+\gamma_j^{(l)}\big)}
\\&\times
\frac{\Upsilon\big(N(\frac{Q}{2}+\beta^{(0)})\big)
\prod_{l=1}^{N}\Upsilon\big(N({\textstyle\frac{Q}{2}}-\mu_l)\big)
\prod_{i<j\leq N}\Upsilon\big(\beta^{(N)}_j-\beta^{(N)}_i\big)}
{\prod_{i\leq N}\Upsilon \big(\frac{Q}{2}+\beta^{(N-1)}-\gamma^{(N-1)}_i\big)
\prod_{i\leq N}\prod_{j\leq N}\Upsilon\big(\frac{Q}{2}-\mu_N+\gamma^{(N-1)}_i-\beta^{(N)}_j\big)}
\eal
The numerator on the second line matches the product of 1-loop determinants of the 
vector multiplets \eqn{vector} with the identification $\gamma^{(l)}_i=a^{(l)}_i$. 
The denominator for $l>1$ is the same as the 1-loop determinant of the bi-fundamental 
fields \eqn{bi-fund}, with the identification
\beq
\frac{N}{l}\beta^{(l-1)}-\frac{N}{l+1}\beta^{(l)}=m_{l-1}
\qquad
l=2,\cdots N-1\,.
\eeq
Note that using the parametrization \eqn{alphai} this can also be written as
\beq
\mu_l-\bar\beta^{(l-1)}+\bar\beta^{(l)}=m_{l-1}\,.
\eeq
The terms in the numerator of the last line depend only on the external states and can be 
removed by field redefinitions. The denominator in the second line for $l=1$ and that in the 
last line are the same as the fundamental and anti-fundamental fields in \eqn{fund} with
\beq
N\beta^{(0)}-\frac{N}{2}\beta^{(1)}=-\hat m_2\,,
\qquad
\beta^{(N-1)}=\hat m_1\,,
\qquad
\mu_N+\beta^{(N)}_i=\tilde m_i\,.
\eeq
Lastly we have the relation \eqn{constraint}
\beq
\beta^{(l-1)}-\beta^{(l)}=\mu_l\, .
\eeq

This gives a matching between the product of three point functions in Toda CFT and the 
one-loop partition function of the gauge theory, up to terms which can be absorbed in 
overall normalizations. 
The expressions \eqn{beta} and \eqn{initial} are the natural generalization of these conditions 
to a general quiver tail.

\section{Discussion}
\label{sec:discuss}

We have generalized the AGT correspondence to the case of 4d $\cN=2$ linear quiver 
theories with quiver tails. They can be described within 2d conformal $A_{N-1}$ Toda field theory, 
where $N-1$ is the rank of the largest gauge group. As proposed in \cite{KMS}, one should 
consider the correlation function of simple punctures on the sphere with two special 
punctures with specific semi-degenerate states. The smaller gauge groups in the tail are 
represented by subspaces of states of Toda CFT which are, or are not degenerate.

These subspaces of states arose as the result of the successive fusion of the degenerate 
state at the special puncture with the states at the simple punctures. As we have shown in 
Section~\ref{sec:2x3} in the case of the $SU(2)\times SU(3)$ quiver, the restriction to this 
subspace arises as a consequence of a Ward identity for the degeneracy condition of the 
special state. The same should be true more generally, with a special state satisfying 
$N-N_1$ degeneracy conditions and therefore there will be this number of conditions 
on the state $\alpha^{(1)}$, reducing the space of states to be $N_1-1$ dimensional. 
For $A_3$ Toda one should be able to 
use the explicit algebra written down in \cite{Kausch:1990bn, Blumenhagen:1990jv} to derive 
these conditions. 

The other way to see the fusion rules is from studying the three point function of generic 
states and its degenerations. The result of Section~\ref{sec:3pt} is that indeed the 
desired subspace arises in the three point function. We did it in full detail for the 
``simple tail'', but it seems to work more generally. Still, it may be that in certain 
special cases there would be extra states allowed in the OPE. Comparing these 
three point functions to the 1-loop partition function in the gauge theory allowed 
us to identify the full map of parameters between the two picture.

In the case of $SU(2)\times SU(3)$ we studied also the contribution to the 
conformal blocks from level 1 states. 
By the AGT duality, the sum over descendants of the Virasoro algebra is 
equal to the contribution of $SU(2)$ instantons and $W_3$ descendants agree with 
$SU(3)$ instantons. In our case we found that when restricting to the one-dimensional 
subspace of $A_2$ Toda primaries, for which the three point function does not vanish, 
the sum over {\em both} descendants of the $W_3$ algebra at level one adds up 
to the same answer as the $SU(2)$ instantons (with an extra polynomial remnant). 
Though we summed over two descendants, we reproduces the same answer that 
one gets from the single descendant at level one of Liouville.

It is compelling to postulate alternative 2d descriptions of quiver tails in addition to the one 
presented in this paper. One possibility is to couple Toda theories of different 
rank, so the $SU(2)\times SU(3)$ would be described by Liouville coupled to $A_2$ Toda.
Clearly the spaces of primaries of these two theories agrees with the 
Coulomb branch parameters of the two groups without the need to restrict to 
a subspace. Likewise, the descendants are known to reproduce the instanton 
partition functions. What is needed is to find a way to couple the two 2d CFTs in 
a consistent way, which will give the desired answer.

It would be interesting to generalize our construction to other linear theories 
which are not conformal, along the lines of \cite{Gaiotto:2009ma}.

With the map we proposed in this paper it is possible now to study observables 
in these conformal theories with quiver tails. One can introduce surface operators 
\cite{Alday:2009fs,Alday:2010vg,Dimofte:2010tz, Maruyoshi:2010iu,
Taki:2010bj,Awata:2010bz,Kozcaz:2010yp,Wyllard:2010rp, Kozcaz:2010af}, 
Wilson loops and 't~Hooft loops \cite{Alday:2009fs,DGTT,DGG,Passerini:2010pr,Gomis:2010kv} 
and domain walls \cite{DGG, Hosomichi:2010vh} 
and see how they behave when coupled to the lower rank gauge groups in the tail.

\section*{Acknowledgements}

We are grateful to Jaume Gomis, Alexey Litvinov, Takuya Okuda, Sara Pasquetti
and Gerard Watts for interesting discussions and the JHEP referee for useful suggestions 
and comments. 
N.D. would like to thank the hospitality of Nordita (Stockholm), the Schr\"odinger Institute (Vienna), 
the KITP (Santa Barbara), POSTECH (Pohang), KIAS (Seoul) and DESY (Hamburg) during the 
course of this work. F.P. would like to thank the Perimeter Institute (Waterloo) for its hospitality. 
This research was supported in part by the National Science Foundation under Grant No. PHY05-51164.
The work of N.D. is underwritten by an advanced fellowship of the 
Science \& Technology Facilities Council.

\appendix

\section{Special Functions }
\label{app:spefunct}

The function $\Gamma_b(x)$ is a close relative of the double
Gamma function studied in \cite{Ba,Sh}. It 
can be defined by means of the integral representation
\beq
\log\Gamma_b(x)=\int_0^{\infty}\frac{dt}{t}
\left(\frac{e^{-xt}-e^{-Qt/2}}{(1-e^{-bt})(1-e^{-t/b})}-
\frac{(Q-2x)^2}{8e^t}-\frac{Q-2x}{t}\right)\,.
\label{barnes}
\eeq
We use the following relation satisfied by this function
\beq
\Gamma_b(x+b)\Gamma_b(x+1/b)=x\,\Gamma_b(x)\Gamma_b(x+b+1/b)\,.
\eeq

The $\Upsilon$ function may be defined in terms of $\Gamma_b$ (with $Q=b+1/b$)
\begin{equation}
\Upsilon(x)\equiv \frac{1}{\Gamma_b(x)\Gamma_b(Q-x)}\,.
\end{equation}
An integral representation convergent in the strip $0<{\rm Re}(x)<Q$ is 
\beq
\text{log}\Upsilon(x)=\int_{0}^{\infty}\frac{dt}{t}\left\lbrack\left(\frac{Q}{2}-x\right)^{2}e^{-t}-\frac{\text{sinh}^{2}(\frac{Q}{2}-x)\frac{t}{2}}{\text{sinh}\frac{bt}{2}\text{sinh}\frac{t}{2b}}\right\rbrack\;.
\eeq
Important properties we need are the obvious reflection $\Upsilon(Q-x)=\Upsilon(x)$ and that 
at integer multiples of $Q$ it has zeros.

\bibliography{refs}
\end{document}